# Towards a Reliable Framework of Uncertainty-based Group Decision Support System

*Junyi CHAI and James N.K. LIU*
Department of Computing,
The Hong Kong Polytechnic University
Hung Hom, Kowloon, Hong Kong, SAR
{ csjchai, csnkliu }@comp.polyu.edu.hk

*Abstract*--This study proposes a framework of Uncertainty-based Group Decision Support System (UGDSS). It provides a platform for multiple criteria decision analysis in six aspects including (1) decision environment, (2) decision problem, (3) decision group, (4) decision conflict, (5) decision schemes and (6) group negotiation. Based on multiple artificial intelligent technologies, this framework provides reliable support for the comprehensive manipulation of applications and advanced decision approaches through the design of an integrated multi-agents architecture.

## I. INTRODUCTION

Nowadays, since companies are usually working in a rapidly changing and uncertain business environment, more timely and accurate information are required for decision-making, in order to improve customer satisfaction, support profitable business analysis, and increase their competitive advantages. In addition to the use of data and mathematical models, some managerial decisions are qualitative in nature and need judgmental knowledge that resides in human experts. Thus, it is necessary to incorporate such knowledge in developing Decision Support System (DSS). A system that integrates knowledge from experts is called a Knowledge-based Decision Support System (KBDSS) or an Intelligent Decision Support System (IDSS) [1]. Moreover, two kinds of situations significantly increase the complexity of decision problem: (1) multiple participants involved in decision process; (2) decision-making under uncertainty environment.

In this paper, we propose a framework of Uncertainty-based Group Decision Support System (UGDSS). Unlike existing DSS designs, this framework is based on multiagent technology and standalone knowledge management process. Through the adoption of agent technologies, this design provides an integrated system platform to resolve the uncertainty problem in support of group Multi-Criteria Decision Making (MCDM). Firstly, we provide a general model of group MCDM. Then, we carry out an analysis on uncertainty-based group MCDM, and present our designing basis of UGDSS. Thirdly, we propose the architectures and structures of UGDSS, including other two kinds of knowledge-related system components: (1) Decision Resource MIS and (2) Knowledge Base Management System (KBMS).

The rest of this paper is organized as follows. Section 2 provides the general problem model of Group MCDM. Section 3 gives the uncertainty analysis of group decision environment. Section 4 presents the framework of UGDSS, including two knowledge-related system components. Section 5 shows our conclusion and outlines for future work.

## II. GROUP MULTIPLE CRITERIA DECISION ANALYSIS

### A. Multiple Criteria Decision Making

Multiple Criteria Decision Making (MCDM) was derived from the Pareto Optimization Concept long time ago. After 50 years, Koopmans [2] introduced the Efficient Point in decision area. At the same time, Kuhn and Tucker [3] introduced the concept of Vector Optimization. Then, Charnes and Cooper [4] studied the model and application of Linear Programming in decision science. In 1972, the International Conference on MCDM held by Cochrane and Zeleny [5] remarked that the normative MCDM theory had been developed as the mainstream of decision science. More recently, many applicable MCDM approaches have been used to design Decision Support System for solving specific domain problems.

The MCDM with certain information and under certain environment is called Classic MCDM. Major methods of Classic MCDM can be roughly divided into three categories: (1) Multiple Criteria Utility Theory, (2) Outranking Relations, (3) Preference Disaggregation.

*1) Multiple Criteria Utility Theory*

Fishburn [6] and Huber [7] provided very specific literature survey on Multiple Criteria Utility Theory. Besides, Keeney and Raiffa [8] published a monograph which deeply influences the future development.

*2) Outranking Relations*

The outranking relations approach aims to compare every couple of alternatives and then gets overall priority ranks, which mainly includes the ELECTRE method and the PROMETHEE method. ELECTRE was firstly proposed by Roy [9] in 1960s. And, PROMETHEE method was initially established by Brans [10]. Xu [11] extended PROMETHEE with a Superiority and Inferiority Ranking (SIR) method which integrated with the outranking approach.

*3) Preference Disaggregation.*

Jacquet-Lagreze et al. [12] provided a UTA method to maximize the approximation of the preference of decision makers by defining a set of additive utility functions. Zopounidis and Doumpos [13][14] developed the UTADIS method as a variant of UTA for sorting problems, and extended the framework of UTADIS for involving multi-participants cases called the MHDIS method.

*B. General Problem Model of Group MCDM*

The group MCDM problem involves multiple participants assessing alternatives based on multiple criteria. In order to facilitate the establishment and development of MCDM support system, we carry out an analysis of group MCDM to form a general model.

Fig. 1 shows our proposed general problem model for group MCDM. It generally contains three decision sets: alternative sets ($Y_i$), decision maker sets ($e_k$) and criterion sets ($G_j$). It involves two kinds of weights: decision maker weights ($w_k$) and criterion weights ($\omega_j$). Decision makers provide their individual decision matrix $d_{ij}^{(k)}$ and the weights ($\omega_j$) of every criterion. In this figure, individual decision information is represented with a decision-plane ($P_k$) including black points ($d_{ij}^{(k)}$) and grey points ($\omega_j^{(k)}$). Therefore, the group aggregation process can be shown as a plane-projection from individual decision plane $P_k$ to group-integrated decision plane $P(K)$

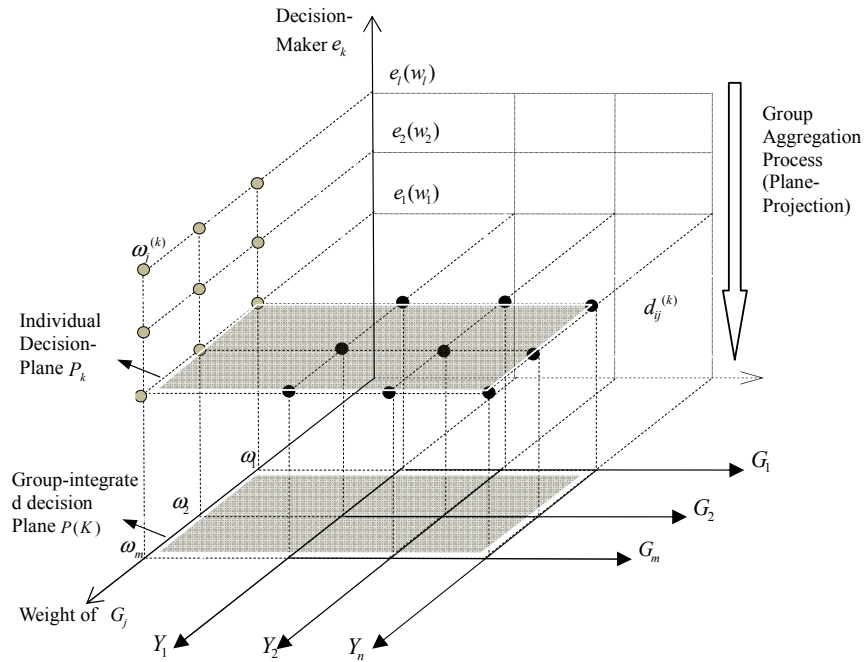

Figure 1. General problem model of Group MCDM

## III. UNCERTAINTY MULTIPLE CRITERIA DECISION ANALYSIS

Although Classic MCDM already has quite a complete theory by now, it still cannot solve most MCDM problems in real world. One main reason is that the decision information are usually not provided completely, clearly or precisely in reality. In most cases, people have to make decisions in uncertainty environment. Therefore, many researchers pay more attention on this research branch of Uncertainty MCDM.

Uncertainty MCDM is non-classic, and can be treated as the extension of Classic MCDM. We can generally divide the uncertainty problems into three categories: (1) Stochastic type (2) Fuzzy type (3) Rough type. A comparison of these uncertainty types is shown in Table I. Accordingly, Uncertainty MCDM also has three research directions: Stochastic MCDM, Fuzzy MCDM and Rough MCDM. Recently, many approaches have been developed to solve these problems (e.g. [15]). This paper will adopt this classification method to establish the framework of UGDSS including its subsystems, several intelligent agents and other functional applets.

TABLE I. The Comparison in Uncertainty Types of Decision Problems

| MCDM | Uncertainty Objects | Research Variable | Major Approaches |
|---|---|---|---|
| Stochastic type | Possibility of decision results is uncertainty (not certain incidence) | decision attributes in decision problem | 1. Utility Theory<br>2. Probability Aggregation<br>3. Stochastic Simulation… |

| Fuzzy type | The membership of objects is uncertainty (not clarity) | The value of decision attribute in decision problem | Fuzzy sets (Intuitionistic Fuzzy, Linguistic Fuzzy, … ) |
| Rough type | The granularity of objects is uncertainty (not accuracy) | Decision schemes | Rough sets (Variable Precision RSs, Dominance Based RSs, … ) |

*1) Stochastic MCDM*

Bayes theory is proposed for stochastic process which can improve the objectivity and veracity in stochastic decision making. Then, Bernoulli [16] introduced the concept of Utility and Expected Utility Hypothesis Model. von Neumann and Morgenstern [17] concluded the Expected Utility Value Theory, proposed the axiomatic of Expected Utility Model, and mathematically proved the results of maximized Expected Utility for decision maker. Wald [18] established the basis of statistical decision problem, and applied them in the selection of stochastical decision schemes. Blackwell and Girshich [19] integrated the subjective probability with the utility theory into a clear process to solve decision problems. Savage [20] extended the Expected Utility Model, and Howard [21] introduced the systematical analysis approach into decision theory and developed them from theory and application aspects.

*2) Fuzzy MCDM*

In 1965, Zadeh [22] proposed the Fuzzy Sets which adopted the membership functions to represent the degree of membership from elements to sets. Atanassov and Gargov [23][24] extended Zadeh's Fuzzy Sets concept into the Intuitionistic Fuzzy Sets (IFSs), and then as in the following, they extended IFSs into the Interval-Valued Intuitionistic Fuzzy Sets (IVIFSs), which are described by a membership degree and a non-membership degree whose values are intervals rather than real numbers. Based on these pioneering works, theories of IFSs and IVIFSs have received much attention from researchers. Until recently, some basic theorems such as Calculation Operators and Fuzzy Measures, have just been founded for various applications[25][26].

In Chai and Liu's earlier work [15], a novel Fuzzy MCDM approach is proposed based on the Intuitionistic Fuzzy Sets (IFSs) theory to solve the real problem in Supply Chain Management (SCM). It firstly applies IFSs to define and represent the fuzzy natural language terms which are used to describe the individual decision values and the weights for decision criteria and decision makers. And then six main steps of this approach are presented to solve uncertainty group MCDM problem. This work enriches the method base of solving fuzzy MCDM problem, and can be implemented in the proposed UGDSS.

*3) Rough MCDM*

Pawlak [27][28] systematically introduced the Rough sets theory. Then, Slowinski [29] concluded the past achievements of Rough sets in theory and applications. Since 1992, the annual International Conference on Rough Sets has been playing a very important role in promoting the development of Rough sets in theory extension and various applications. More recently, Greco [30] proposed a Dominance based Rough Sets theory which produces the decision rules with stronger applicability. By now, Rough Sets theory has been applied in decision analysis, process control, knowledge discovery, machine learning, pattern recognition, etc. In UGDSS, the rough MCDM approaches are mainly implemented by various subsystems with specific function modules, key intelligent agents, and other deployable applets.

IV. UGDSS FRAMEWORK

In this section, we propose the framework of UGDSS. Here, the term "Knowledge" is a comprehensive concept, which includes data, model, human knowledge and other forms of information, so long as it can be used in uncertainty group decision making.

*A. Uncertainty Group Decision Process and System Structure*

In uncertainty group decision process, we mainly consider three factors which increase the complexity of decision-making in reality: (1) Uncertain decision environment, (2) Unstructured decision problem, (3) Complex decision group, and another issue: Group unification of decision conflict. This process provides a mechanism to address the three kinds of complexities and group conflict, which consist of six analysis stages:
- Decision Environment Analysis
- Decision Problem Analysis
- Decision Group Analysis
- Decision Scheme Analysis
- Decision Conflict Analysis
- Group Coordination and Decision Analysis

From decision-makers' view, these stages have the basic logical sequence. Suppose there is a MCDM problem with complex internal structure involving multiple participants. We firstly need to analyze the existing internal and external environments, and figure out what are the decision conditions; whether the decision information is complete, certain and quantizable; what kinds of uncertainty type it belongs to. Secondly, the specific decision problems need to be analyzed, including ontological investigation, problem representation and decomposition, etc. Thirdly, an ontological group analysis is required to reduce the complexity of human organizational structure. Fourthly, people need to establish problem-solving solutions which may be derived from various resources including previous problem-solving schemes in knowledge bases, decision schemes from domain experts, or results of group discussion, etc. Fifthly, we need to integrate those dispersive, multipurpose, individual or incomplete decision opinions into one or a set of applicable final decision results. Besides, the six stages mentioned above can momentarily call Negotiation Support System in conflict analysis stage for possible decision conflicts.

From the view of system process, each stage consists of several subsystems with different functions. For example, we adopt the ontological problem analysis tools to represent, scrutinize and decompose the complex decision problem. These subsystems are integrated in the UGDSS platform with supports of interface technologies and intelligent agent technologies. In this design, parallel computation in subsystems and middleware is quite important, which can produce better system efficiency.

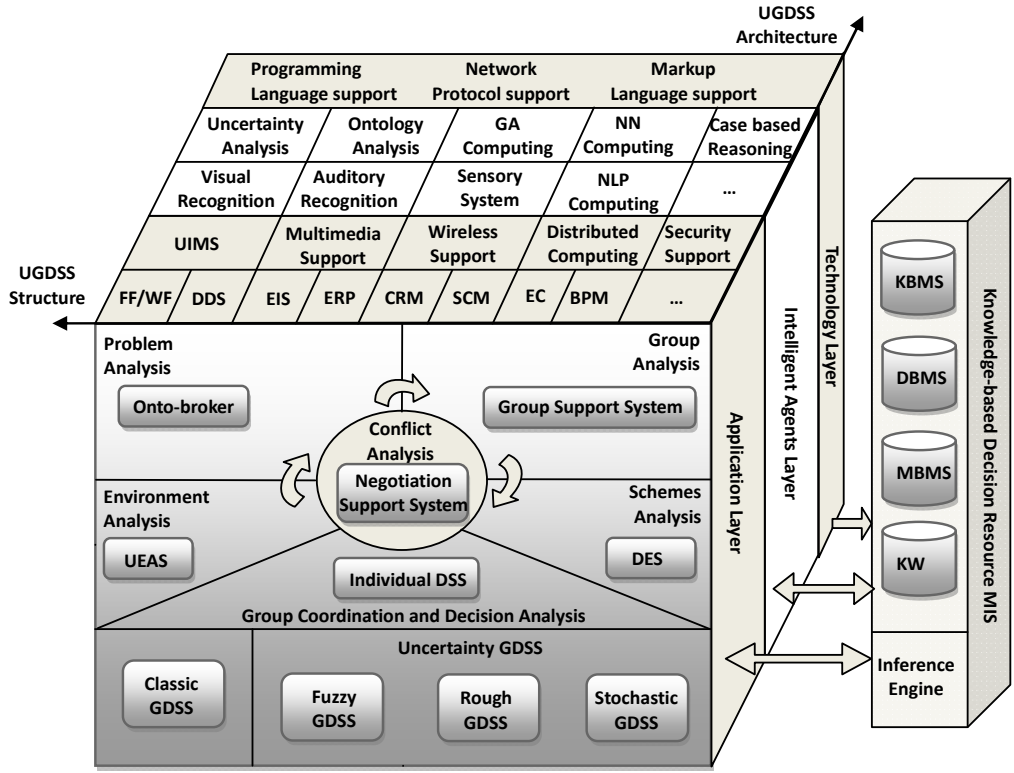

Figure 2. UGDSS Architecture

*1) Decision Environment Analysis*

Decision environment is an important factor which significantly influences other decision stages. It may contain different aspects such as decision targets, decision principles, possible limitations, available resources, etc. More importantly, people need to analyze whether there are any uncertainty information. In this paper, we define that the uncertainty decision information consists of the following situations:

- Information deficiency
- Information incompletion
- Dynamic information
- Unclarity information
- Inaccuracy information
- Multiple uncertainties

Although several MCDM approaches have been developed, it is not enough for solving complex uncertainty decision problem in reality. Therefore, one of our future works aims to establish an Uncertainty Environment Analysis Sub-system (UEAS) to handle uncertainty information, and then extend its capability to solve other uncertainty MCDM problems.

*2) Decision Problem Analysis*

We can generally divide decision problems into three categories: (1) Structure, (2) Semi-structure, (3) Non-structure. To the first one, problems are well organized and represented for ontological analysis and decomposition. To another two, problems are usually represented in the form of text or interviewing dialogues. Therefore, these problems need to be ontologically represented and described at this stage. Some useful analysis technologies can be adopted including ontological analysis in Onto-broker [31], natural language process, etc.

*3) Decision Group Analysis*

Many decision problems in reality (such as great strategic decision of government or industry, the organizational decision of large corporation, etc.), involve multiple participants with complex human relationship or organizational structure. A good group analysis can result in much efficient decision process and impartial decision results. Group Support System (GSS) is used for group analysis including decomposition, reorganization, character analysis, integration, etc. Some methods such as Double Selection Model [32] are a feasible approach to realize group analysis in GSS.

*4) Decision Scheme Analysis*

Decision schemes are the problem-solving solutions to specific decision problem. These schemes may be derived from previous decision schemes reorganized in Scheme Base; new problem-solving schemes established by domain experts; solutions produced in group discussion and negotiation; all kinds of information on Web or somewhere, etc. This stage is supported by Domain Expert System and corresponding Decision Resource MIS.

*5) Decision Conflict Analysis*

Decision conflict analysis is the core process in UGDSS. The conflicts may be derived at each stage of decision-making process. Therefore, the subsystem at each stage may call the programs of Negotiation Support System (NSS) for conflict analysis. Chai and Liu's earlier work [32] provided a Group Argumentation Model in order to solve complex decision conflicts. This model can be used to design and develop Negotiation Support System.

*6) Group Coordination and Decision Analysis*

This stage takes the responsibility for the integration of group opinion. Many methods can be used to solve this problem like Vector Space Clustering, Entropy Weight Clustering, Intuitionistic Fuzzy Weight Average (IFWA) method [26], Weighted Group Projection method [15], etc. Besides, Individual Decision Support System is a helper of decision maker to develop their own opinion, corresponding with Domain Expert Systems to form high quality individual decision schemes.

B. *UGDSS Architecture*

Unlike existing designs of DSS which mainly focus on specific problem domains, the UGDSS architecture provides an integrated system platform for complete decision analyses and comprehensive applications. The system architecture is shown in Fig. 2, which consists of three layers:
1. The Application Layer
2. The Intelligent Agents Layer
3. The Technology Layer

*1) Application Layer*

*a) Basic Function Modules*
- User Interface Management System (UIMS)
  UIMS, as a subsystem of UGDSS, is composed of several programs and functional interface components in intelligent agent layer such as natural language process, uncertainty analysis process, visual reorganization function, etc.
- Multimedia support
  Multimedia technologies are comprehensively used in UGDSS. The interfaces in application layer are related to many intelligent agents including Visual recognition, Audio recognition, etc.
- Wireless support
  Many mobile application devices such as PDA, mobile phone, wireless facilities are used to support group decision-making
- Security support
  In order to guarantee the security of system and data transmission, security support is indispensible in system establishment. Some main technologies like internal control mechanism, firewall, ID authentication, encryption techniques, digital signature, etc.

*b) Application Domain Modules*

The application domain modules aim to solve specific problems in different domains. For example, Chai and Liu's Fuzzy MCDM method [15] is used to solve the problem of supply chain partner selection. This application requires general domain knowledge of Supply Chain Management. These application modules as middleware of UGDSS can provide the necessary supports to various specific application domains. Several domains are given in following.
- Financial/Weather Forecasting (FF/WF)
- Director Decision Support (DDS)
- Enterprise Information System (EIS)
- Enterprise Resource Planning (ERP)
- Customer Relationship Management (CRM)
- Supply Chain Management (SCM)
- E-Commerce (EC)
- Business Process Management (BPM)

*2) Intelligent Agent Layer*
- Sensory system
  Sensory systems, such as vision systems, tactile system, and signal-processing systems, provide a tool to interpret and analyze the collected knowledge and to respond and adapt to changes when facing different environment.
- Genetic Algorithm (GA) computing agent
  Genetic Algorithms are sets of computational procedures, which learnt by producing offspring that are better and better as measured by a fitness function. Algorithms of this type have been used in decision-making process such as Web search, financial forecasting, vehicle routing, etc.
- Neural Network (NN) computing agent
  A Neural Network is a set of mathematical models which simulate the way a human brain functions. A typical intelligent agent based on NN technology can be used in stock forecasting for decision making.
- Uncertainty analysis agent
  This agent is used to analyze the environment and conditions of decision problem.
- Case Based Reasoning (CBR) agent
  Case Based Reasoning is a means for solving new problems by using or adapting solutions of old problems. It provides a foundation for reasoning, remembering, and learning. Besides, it simulates natural language expressions, and provides access to organizational memory.
- Natural Language Process (NPL) computing

Natural Language Process (NLP) technology provides people the ability to communicate with a computer in their native language. The goal of NLP is to capture the meaning of sentences, which involves finding a representation for the sentences that can be connected to more general knowledge for decision making.

Besides, all of these intelligent agents with various group decision functions may consist of different kinds of knowledge/information bases which are united and embodied in Decision Resource Management Information System (DRMIS). This design can improve the efficiency of information processing and the robustness of system.

*3) Technology Layer*

This layer provides the necessary system supports to other two layers and DRMIS. It mainly includes (1) programming language support (VS.Net, C#, Java, etc.) (2) network protocol support (HTTP, HTTPS, ATP, etc.) (3) markup language support (HTML, XML, WML, etc). Besides, the technology layer also provides various technology supports for constructing the four Bases, Inference engine, etc.

*C. Knowledge-related System Designs*

*1) Knowledge-based Decision Resource MIS framework*

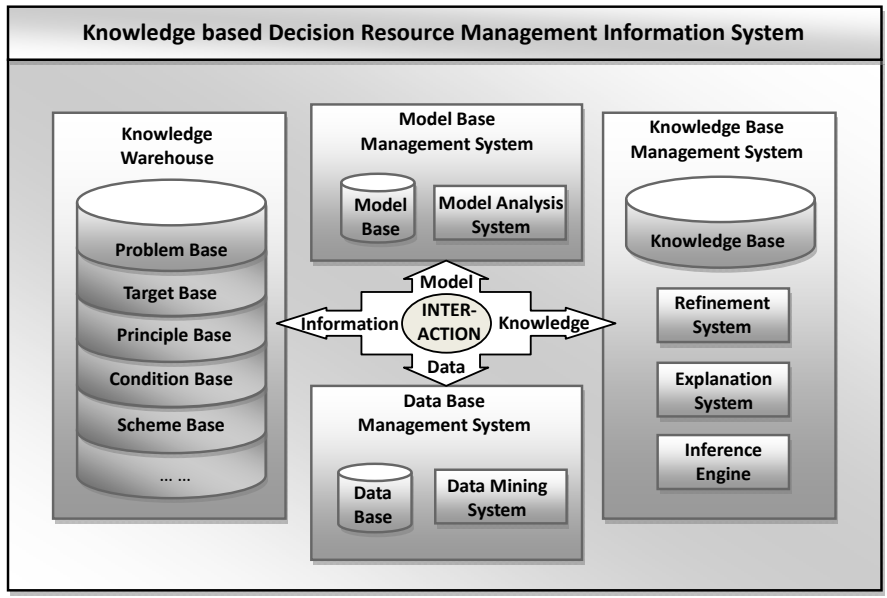

Figure 3. Knowledge-based Decision Resource MIS Framework.

Fig. 3 shows the framework of Knowledge-based Decision Resource MIS. It mainly consists of four kinds of subsystem: KBMS, DBMS, MBMS, KW. In this system, different kinds of information, knowledge, models and data interact together and provide the supports to the whole UGDSS.

- *Data Base Management System (DBMS)*

Generally, DSS needs a standalone database. Especially, this DSS is required to solve the complex uncertainty group decision problems. Therefore, DBMS is a necessary component in UGDSS, which consists of a DSS database and a Data Mining System. A database is created, accessed, and updated by a DBMS. And Data Mining System is used to discover knowledge from data resources. Many technologies are applicable to mining data, such as statistical approaches (Bayes's theorem, cluster analysis, etc), case-based reasoning, neural computing, genetic algorithms, etc. These technologies are developed as intelligent agents located in the second layer in Fig. 2.

- *Model Base Management System (MBMS)*

MBMS mainly includes Model Base and Model Analysis System. Model base contains routines and special statistics, financial forecasting, management science, and other quantitative models which provide the resources for Model Analysis System. Turban [1] divided the models into four major categories: Strategic, Tactical, Operational, and Analytical. In addition, there are model building blocks and routines. Based on these model resources, Model Analysis System is used to build blocks; generate the new routines and reports; update and change model; and manipulate model data, etc.

- *Knowledge Base Management System (KBMS)*

There are three kinds of knowledge which will be used in decision-making: (1) structure (2) semi-structure (3) non-structure. The structural knowledge is usually reorganized in available models and stored in model base. Much semi-structural and non-structural knowledge are so complex that they cannot be easily represented and reorganized. Therefore, more professional knowledge processing system called KBMS is required to enhance the capability of knowledge management. In next section, we present a detailed knowledge management process in

KBMS.

● *Knowledge Warehouse (KW)*

In UGDSS, KW mainly contains multiple bases for classified storage of decision knowledge including decision problems, targets, principles, conditions, schemes, etc. It is responsible for storage, extraction, maintenance, interaction and other knowledge manipulations.

2) *Knowledge Management Process in KBMS*

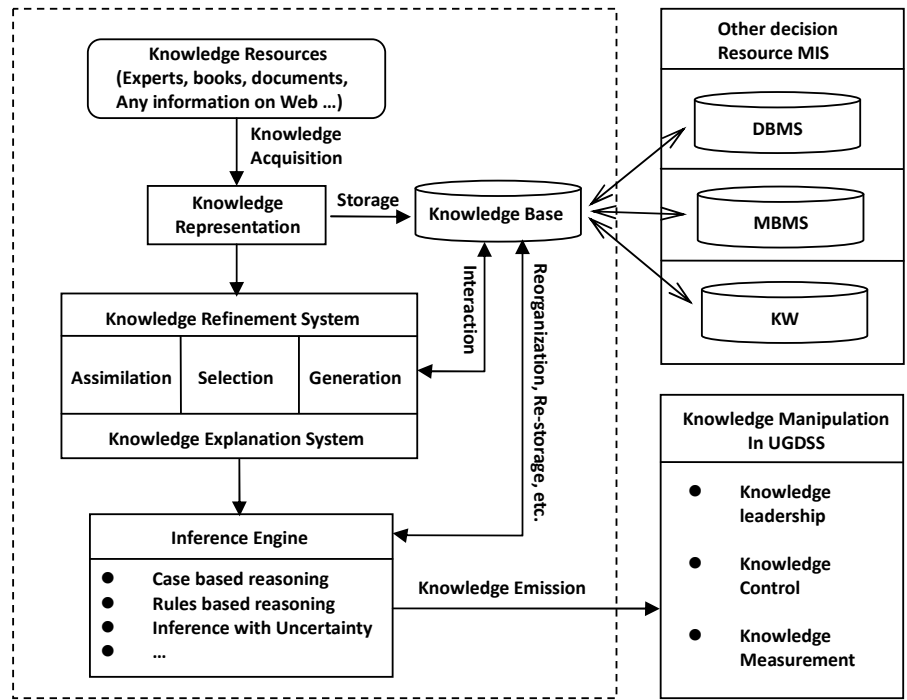

Figure 4. Knowledge Management Process in KBMS

In this process, there are five basic classes of knowledge manipulation activities including: acquisition, selection, generation, assimilation, and emission [33]. These activities are the basis for problem founding and solving, as well as being involved at each stage in the decision making process. In this paper, we provide a knowledge management process in KBMS (see Fig. 4).

● *Knowledge Resource:* Some possible knowledge sources include domain experts, books, documents, computer files, research reports, database, sensors, and any information available on the Web.

● *Knowledge Acquisition:* This activity is the accumulation, transmission, and transformation of documented knowledge resources or problem-solving scheme of experts.

● *Knowledge Representation：* The acquired knowledge is organized in this activity, which involves preparation of a knowledge map and encoding the knowledge in the knowledge base.

● *Knowledge Selection：* In knowledge refinement and explanation system, the knowledge is validated and verified until its quality is acceptable. There are three activities to refine and explain the acquired knowledge: (1) selection, (2) generation, (3) assimilation. In selection activity, systems select knowledge from information resources and making it suitable for subsequent use.

● *Knowledge Generation：* In this activity, knowledge is produced based on the decision incident by either discovery or derivation from existing knowledge.

● *Knowledge Assimilation：* In assimilation activity, this knowledge refinement and explanation system alter the state of the decision makers' knowledge resources by distributing and storing the acquired, selected, or generated knowledge [33].

● *Inference Engine：* In knowledge base, knowledge has been organized properly and represented in a machine-understandable format. The inference engine can then use the knowledge to infer new conclusions from existing facts and rules. There are many different ways of representing human knowledge, including Production rules, Semantic networks, Logic statement, and Uncertainty information representation, etc. Here, knowledge is recognized and restored in knowledge base which also conducts the communication with other DRMISs.

● *Knowledge Emission：* This activity embeds knowledge into the outputs of KBMS, and input the useful knowledge of specific decisional episode into UGDSS for further decisional knowledge manipulation activities including knowledge leadership, control and measurement.

## V. CONCLUSION

This paper proposes the framework of Uncertainty Group Decision Support System (UGDSS) and other two kinds of knowledge-related system components: Decision Resource MIS and Knowledge Base Management System (KBMS). We firstly provide a general problem model of group MCDM. And then, we carry out an analysis on uncertainty group MCDM problem, and present the basis of system design on handling the uncertainty decision-making. Finally, we propose a set of detailed designs of system architectures and structures for supporting a complete uncertainty group decision-making process including (1) environment analysis, (2) problem analysis, (3) group analysis, (4) scheme analysis, (5) group coordination analysis and (6) decision conflict analysis.

In future, we will make effort on two directions. In system aspect, we need to develop multiple intelligent agents, middleware or subsystems which are integrated in UGDSS. In decision theory aspect, we will consider how to develop more applicable uncertainty group decision-making approaches based on Fuzzy sets, Rough sets, Grey system theory and other uncertainty theories.

## ACKNOWLEDGMENT

The authors would like to acknowledge the partial supports from the GRF 5237/08E of the Hong Kong Polytechnic University.